\documentclass[8.5pt,twoside,twocolumn]{article}
\usepackage[super,sort&compress,comma]{natbib} 
\usepackage[version=3]{mhchem}
\usepackage{balance}
\usepackage{times,mathptm}
\usepackage{sectsty}
\usepackage{graphicx} 
\usepackage{lastpage}
\usepackage[format=plain,justification=raggedright,singlelinecheck=false,font=small,labelfont=bf,labelsep=space]{caption} 
\usepackage{fancyhdr}
\usepackage{amssymb}

\begin{document}

\thispagestyle{plain}
\fancypagestyle{plain}{
\fancyhead[L]{\includegraphics[height=8pt]{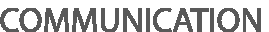}}
\fancyhead[C]{\hspace{-1cm}\includegraphics[height=20pt]{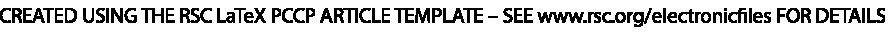}}
\fancyhead[R]{\hspace{10cm}\vspace{-0.25cm}\includegraphics[height=10pt]{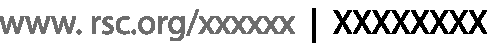}}
\renewcommand{\headrulewidth}{1pt}}
\renewcommand{\thefootnote}{\fnsymbol{footnote}}
\renewcommand\footnoterule{\vspace*{1pt}%
\hrule width 3.4in height 0.4pt \vspace*{5pt}} 
\setcounter{secnumdepth}{5}

\makeatletter 
\renewcommand\@biblabel[1]{#1}            
\renewcommand\@makefntext[1]%
{\noindent\makebox[0pt][r]{\@thefnmark\,}#1}
\makeatother 
\renewcommand{\figurename}{\small{Fig.}~}
\sectionfont{\large}
\subsectionfont{\normalsize} 

\fancyfoot{}
\fancyfoot[LO,RE]{\vspace{-7pt}\includegraphics[height=9pt]{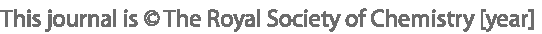}}
\fancyfoot[CO]{\vspace{-7.2pt}\hspace{12.2cm}\includegraphics{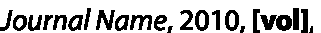}}
\fancyfoot[CE]{\vspace{-7.5pt}\hspace{-13.5cm}\includegraphics{headers/RF}}
\fancyfoot[RO]{\footnotesize{\sffamily{1--\pageref{LastPage} ~\textbar  \hspace{2pt}\thepage}}}
\fancyfoot[LE]{\footnotesize{\sffamily{\thepage~\textbar\hspace{3.45cm} 1--\pageref{LastPage}}}}
\fancyhead{}
\renewcommand{\headrulewidth}{1pt} 
\renewcommand{\footrulewidth}{1pt}
\setlength{\arrayrulewidth}{1pt}
\setlength{\columnsep}{6.5mm}
\setlength{\bibsep}{1pt}

\newcommand{\dtwomin}{D^2_\mathrm{min}}

\twocolumn[
  \begin{@twocolumnfalse}
\noindent\LARGE{\textbf{Yielding and microstructure in a 2D jammed material under shear deformation$^\dag$}}
\vspace{0.6cm}

\noindent\large{\textbf{Nathan C.\ Keim\textit{$^{a}$} and
Paulo E.\ Arratia$^{\ast}$\textit{$^{a}$}}}\vspace{0.5cm}

\noindent\textit{\small{\textbf{Received Xth XXXXXXXXXX 20XX, Accepted Xth XXXXXXXXX 20XX\newline
First published on the web Xth XXXXXXXXXX 200X}}}

\noindent \textbf{\small{DOI: 10.1039/b000000x}}
 \end{@twocolumnfalse} \vspace{0.6cm}

  ]

\noindent\textbf{The question of how a disordered material's microstructure translates into macroscopic mechanical response is central to understanding and designing materials like pastes, foams and metallic glasses. Here, we examine a 2D soft jammed material under cyclic shear, imaging the structure of $\mathbf{\sim 5 \times 10^4}$ particles. Below a certain strain amplitude, the structure becomes conserved at long times, while above, it continually rearranges. We identify the boundary between these regimes as a yield strain, defined without rheological measurement. Its value is consistent with a simultaneous but independent measurement of yielding by stress-controlled bulk rheometry. While there are virtually no irreversible rearrangements in the steady state below yielding, we find a largely stable population of plastic rearrangements that are reversed with each cycle. These results point to a microscopic view of mechanical properties under cyclic deformation.}

\section*{}
\vspace{-1cm}
\footnotetext{\dag~Electronic Supplementary Information (ESI) available: [details of any supplementary information available should be included here]. See DOI: 10.1039/b000000x/}


\footnotetext{\textit{$^{a}$~Dept.\ of Mechanical Engineering and Applied Mechanics, University of Pennsylvania, 220 South 33rd St., Philadelphia PA 19104, USA; E-mail: parratia@seas.upenn.edu}}

Disordered materials such as pastes and concentrated emulsions, in which each particle is constrained by its neighbors, may be formally described as jammed~\cite{VanHecke:2010go}. The inhomogeneous, unsteady way that such materials deform is common across length scales from molecules to emulsion droplets to foam bubbles, yet it seems exquisitely dependent on a sample's particular microstructure and history~\cite{Gopal:1995bg, Falk:1998wm, Schall:2007fd, Manning:2011ha}. This dependence hinders detailed experimental observations and useful models that can relate the arrangement of constituent atoms or particles to bulk stiffness, plasticity, failure, and so on. Such difficulty is especially evident at the yield stress or strain, below which the material is approximately a solid, and above which it flows~\cite{Chen:2010jn}. Yielding is a crucial aspect of a material's bulk behavior, but in rheological measurements the yielding transition can be difficult to pinpoint, and may depend on material history and details of the test~\cite{Stokes:2004fz,Bonn:2009gd,Moller:2009fy}. Uncertainty about both the microscopic mechanisms at work as a material yields, and the definition of yield stress and strain, would be addressed by a microstructural perspective on bulk yielding. With such a perspective, it may be possible to define an intrinsic yield stress (and strain) for a generic material, as well as to design materials to control their plastic behavior.

\begin{figure}
\begin{center}
\includegraphics[width=2.7in]{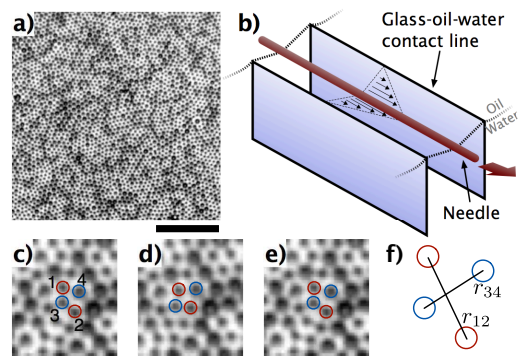}
\end{center}
\caption{ Experimental system. 
\textbf{(a)} Monolayer of particles at oil-water interface, viewed from above. Scale bar is 100~$\mu$m.
\textbf{(b)} Apparatus: A steel needle is adsorbed at the interface, between two glass walls. Magnetic forcing of the needle produces uniform shear of the monolayer (velocity profile shown).
\textbf{(c--e)} Changes to microstructure are observed as T1 events, wherein particles switch neighbors. Sequence shows reversible event over 1 cycle of deformation.
\textbf{(f)} Relative change in $r_{12}$ and $r_{34}$ is used to quantify significance of T1 events.
\label{fig:apparatus}
}
\end{figure}

The relationship between microstructure and yielding is encountered in a simple experimental test: whether microstructure is changed by deforming the whole material to a strain amplitude $\gamma_0$, then reversing that deformation. In the limit $\gamma_0 \to 0$ (far below yielding) the microstructure is unchanged by this procedure if thermal motion is negligible, while for $\gamma_0 \to \infty$ (far above yielding), the original arrangement of particles is completely and irreversibly lost. The transition between these extremes is less straightforward, both in microscopic behavior --- such as whether changes to microstructure are permanent~\cite{haw98pre,corte08,Slotterback:2012fa} --- and bulk rheology~\cite{Stokes:2004fz,Bonn:2009gd,Chen:2010jn}. Several experimental studies have probed the possible connection between reversibility of microstructure and bulk yielding. H\'ebraud et al.~\cite{Hebraud:1997ef} used diffusing-wave spectroscopy to stroboscopically measure changes to a jammed emulsion under cyclic shear, showing that below the rheological yield strain (measured separately), up to several percent of particles were rearranging irreversibly with each cycle. Similar diffusing-wave studies of hard-sphere colloidal glasses~\cite{Petekidis:2002km,Petekidis:2002by} showed instead that yielding marked the division between completely reversible and irreversible behaviors. These differing results were both in the steady state and did not consider transient behavior. 

Here, we examine a 2-dimensional soft jammed material, a disordered suspension of colloidal particles at an oil-water interface on which we can perform bulk rheometry while tracking individual particle rearrangements. Particle tracking leads us to a microscopically-based definition of oscillatory yield strain $\gamma_y^\mathrm{micro}$ and stress $\sigma_y^\mathrm{micro}$ for the material, which is consistent with the corresponding quantities $\gamma_y^\mathrm{rheo}$ and $\sigma_y^\mathrm{rheo}$ measured \emph{simultaneously} by stress-controlled shear rheometry. At stresses below $\sigma_y^\mathrm{micro}$, irreversibility of the microstructure is transient, while above $\sigma_y^\mathrm{micro}$, microstructure changes continually, even after many cycles. The system's ability to eventually reach a reversible steady state thus defines a global yield stress and strain, independent of but consistent with bulk rheology. Consequently, the many \emph{reversible} rearrangements in our experiments are local microscopic plastic events, but are not associated with bulk rheological yielding.

We use a bidisperse mixture of 4.1 and 5.6 $\mu$m-diameter particles adsorbed at an oil-water interface.$^\dag$ The particles have dipole-dipole repulsion~\cite{Aveyard:2000io,Park:2008fi} and so form a stable disordered jammed material (Fig.~\ref{fig:apparatus}a) in which we can continually image and track nearly all particles in a selected region. Rheometry suggests a jamming transition at $\phi \lesssim 0.36$; experiments discussed here are performed at $\phi \sim 0.43$. Brownian motion is not observed and is negligible: the ratio of the diffusion timescale to the bulk deformation timescale, the oscillatory P\'eclet number, is $Pe = \omega R^2 (6\pi \eta R / kT) \gtrsim 10^2$, where $\omega$ is the angular frequency of driving, $\eta$ is solvent viscosity, and $R$ is particle radius. The material is subjected to a linear shear deformation generated by an interfacial stress rheometer (ISR)~\cite{Brooks:1999ky,Reynaert:2008dm}$^\dag$. As shown in Fig.~\ref{fig:apparatus}b, a magnetized needle (diameter 0.23~mm) is placed on the interfacial material to be studied, inside an open channel of width 3.5~mm and length 18~mm formed by 2 walls. Electromagnets move the needle back and forth, shearing the interface in the channel. When the needle's response is measured for a known forcing, the device functions as a sensitive stress-controlled rheometer. 

To consistently prepare the material for each experiment, oscillatory forcing at large amplitude ($\gamma_0 \sim 0.5$) is performed for 6 cycles and then stopped, upon which the system comes to rest with a residual strain $\gamma(t=0) \sim 0.03$, and hence an unrelaxed residual stress $\sigma_\mathrm{DC}(t=0) \sim 15$~nN/m. Thereafter, negligible relaxation is observed on a timescale of 60~s, and waiting times of 1--5 minutes seem to have no effect on subsequent behavior. The material relaxes when we restart oscillatory shear at much smaller amplitude. Such mechanically-activated aging is consistent with generic models of soft glassy material~\cite{Sollich:1997cp, Hebraud:1998gj, Fielding:2009js} and the effect of oscillatory shear on the relaxation time of soft colloidal glasses~\cite{Viasnoff:2002dx}. 

We observe relaxation during shear as a series of rearrangements of the particles. To quantify change to microstructure between two instants, we compare the nearest-neighbor relationships among particles, and then count T1 rearrangements, illustrated in Fig.~\ref{fig:apparatus}(c--e).$^\dag$ T1 rearrangements are a robust way to discretize small changes to microstructure. We remove spurious events due to positional noise by setting a significance threshold for change in $r_{12}$ and $r_{34}$ (Fig.~\ref{fig:apparatus}f)$^\dag$. 

\begin{figure}[h]
\begin{center}
\includegraphics[width=2.5in]{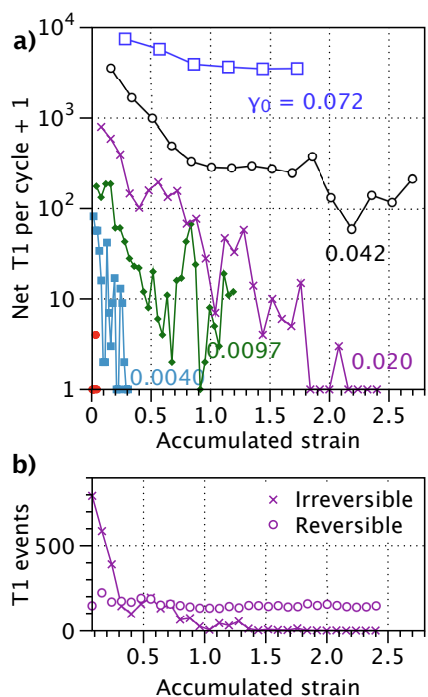}
\end{center}
\caption{Relaxation under cyclic shear, following preparation with unrelaxed stress, plotted as function of accumulated strain ($4 N \gamma_0$ where $N$ is number of cycles). 
\textbf{(a)} Net change in microstructure after each cycle, measured by counting T1 rearrangements. Curves are labeled by strain amplitude; the unlabeled smallest curve is $1.1 \times 10^{-3}$. Below the yield strain, the system relaxes to steady state with conserved microstructure. 
\textbf{(b)} Number of irreversible and reversible T1 events generated in each cycle at $\gamma_0 = 0.020$. The population of reversible rearrangements is stable even as irreversible ones are depleted.
\label{fig:relaxation}
}
\end{figure}

Figure~\ref{fig:relaxation}a shows the net number of irreversible T1 events generated by each cycle of driving plotted as a function of accumulated strain, $\gamma_\mathrm{acc} = 4 N \gamma_0$ where $N$ is the number of cycles. At larger strain amplitudes, after an initial transient, each cycle generates a roughly constant number of new rearrangements. At small amplitudes, however, the net result of the irreversible rearrangements is that the system evolves to a reversible state. The number of new rearrangements eventually decays to zero, out of a total $5 \times 10^4$ tracked particles. Whether oscillatory driving begins by adding to or subtracting from $\sigma_\mathrm{DC}$ has little effect past the first 1--2 cycles.

Our system thus shows 2 regimes of steady-state microstructure behavior, separated by some strain amplitude $\gamma_y^\mathrm{micro}$, which we can define as the largest $\gamma_0$ for which the system can attain reversibility. However, even as the rate of irreversible rearrangements approaches zero, we find that the rate of \emph{reversible} rearrangements --- activated and reversed before the completion of each cycle --- remains nearly constant (Fig.~\ref{fig:relaxation}b). Reversible events may be divided into 2 populations: $\sim 35\%$ do not repeat in subsequent cycles, while $\sim 65\%$ are activated in many consecutive cycles. Thus there remains a large, stable population of reversible plastic rearrangements that play a role in this cyclic deformation, which at the bulk scale is primarily elastic.

\begin{figure}[h]
\begin{center}
\includegraphics[width=2.5in]{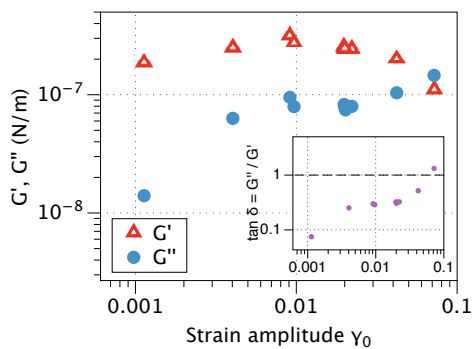}
\end{center}
\caption{Bulk rheology of the interfacial material shown in Fig.~\ref{fig:apparatus}a at $\phi \sim 0.43$. $G'$ and $G''$ are the elastic and loss moduli$^\dag$. Response shows near-linear elasticity at small strain amplitude, but a yielding transition and loss of rigidity at $\gamma_0 \gtrsim 0.03$ ($\sigma_\mathrm{AC} \gtrsim 9$ nN/m). Inset: Ratio $G'' / G'$ as function of $\gamma_0$. 
\label{fig:rheology}
}
\end{figure}

Figure \ref{fig:rheology} shows the steady-state oscillatory rheology of the bulk material measured \emph{simultaneously} in the same experiments, as a function of strain amplitude $\gamma_0$; the imposed stress amplitude $\sigma_\mathrm{AC}$ is varied from 0.52 to 15~nN/m. All measurements are at 0.1 or 0.2~Hz; we have observed that under these experimental conditions, oscillatory rheology varies little with frequency. Measurements reflect nearly-uniform shear deformation of the material in the steady state$^\dag$; under no conditions, including shearing with $\gamma_0 \sim 0.5$, do we observe a persistent localization of strain rate (a ``shear band''). 

The rheometry in Fig.~\ref{fig:rheology} shows that, for a wide range of strain amplitudes, the material is primarily elastic ($G' > G''$), but at a strain amplitude $\gamma_0 \sim 0.03$ it begins to lose rigidity. Nearly-constant elastic and loss moduli giving way to flow at large amplitudes is typical of soft glassy materials~\cite{Spaepen:1977jw, MASON:1995vt, Sollich:1997cp, Hebraud:1998gj, Chen:2010jn, VanHecke:2010go}. Here, we identify this transition with a rheological yield strain amplitude $0.020 \le \gamma_y^\mathrm{rheo} \le 0.042$ and stress amplitude $5.8 \le \sigma_y^\mathrm{rheo} \le 11.7$~nN/m.

\begin{figure}
\begin{center}
\includegraphics[width=3.1in]{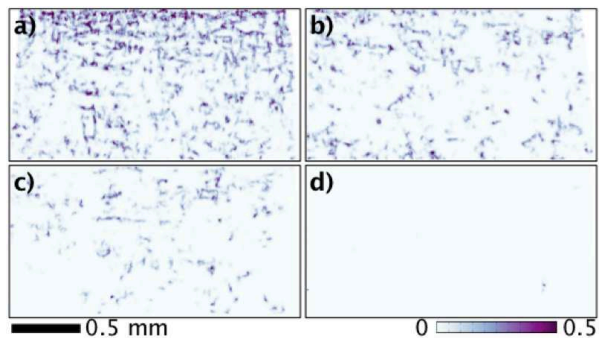}
\end{center}
\caption{
Localization of rearrangements under deformation, above and below yielding. Local non-affinity $\dtwomin$ (see text) is shown for net particle displacements over \textbf{(a)} the first cycle of deformation with $\gamma_0 = 0.07$, \textbf{(b)} 7th cycle; \textbf{(c)} cycles 1--3 at $\gamma_0 = 0.02$ [comprising similar $\gamma_\mathrm{acc}$ as (a, b)], \textbf{(d)} cycles 10--12. The moving needle is at the top of each plot; the wall is at the bottom. High $\dtwomin$ indicates rearrangement activity, occurring in clusters throughout the material.
\label{fig:localization}
}
\end{figure}

We may also ask how relaxation activity is distributed within the portion of the material we observe. This is revealed by the local non-affinity $\dtwomin$ of the particle displacements. We consider each particle and its neighbors within a radius $2.5a$ at time $t$, and computing the affine transformation that best relates their present arrangement to that at $t-\Delta t$, where $a$ is the mean nearest-neighbor spacing. $\dtwomin$ is then the sum of the squared displacements that \emph{cannot} be accounted for by this transformation, and was identified by Falk and Langer~\cite{Falk:1998wm} as indicating particle rearrangements; it is normalized by the number of neighbors considered and by $a^2$. The resulting field is plotted in Fig.~\ref{fig:localization} at early and late times above and below $\gamma_y^\mathrm{micro}$. When significant activity is present, it is distributed throughout the material, with some concentration near the needle in Fig.~\ref{fig:localization}a presumably due to a slight non-uniformity in conditions during large-amplitude preparation. Activity in $\dtwomin$ at late times is consistent with Fig.~\ref{fig:relaxation}$^\dag$.

Our experiments are by far the strongest evidence that a jammed material may approach completely reversible microstructure under cyclic forcing. We note that reversibility does not exclude rearrangements during shear, which we observe, but means only that virtually all rearrangements are reversed at the end of each cycle. This transition was hinted at by the work of Lundberg et al.~\cite{lundberg08} for another soft, frictionless jammed system (2D foam) which was shown to have both reversible rearrangements, and irreversible ones that diminished during the first 2 shear cycles, but those experiments were above yield strain and did not show a transition to complete reversibility as we observe. The experiments of Slotterback et al.~\cite{Slotterback:2012fa} for a jammed frictional packing showed increased reversibility after many cycles at small $\gamma_0$, but here, we have demonstrated that a soft frictionless system may be expected to attain exact reversibility. It is likely that the reversible steady state formed by a particular $\sigma_\mathrm{AC}$ encodes a ``memory'' of that amplitude that may be read and manipulated~\cite{Keim:2011dv}.

Even when the system is unyielded and its microstructure has become reversible, there exist a large number of reversible rearrangements. This can be considered an illustration of shear transformation zones (STZs)~\cite{Falk:1998wm}, microscopic groups of $\mathcal{O}(10)$ particles which undergo a reversible two-state rearrangement under stress. Their hysteretic nature is consistent with our observation that in the steady state at $\gamma_0 = 0.020$, T1 events activate and deactivate at different strains, so that $\langle \gamma_\mathrm{on} - \gamma_\mathrm{off} \rangle = 0.014 \pm 0.001$ (std.\ deviation of mean). STZs are thought to be elemental loci of plasticity, and have informed models of deformation in soft glassy matter~\cite{Falk:1998wm, Schall:2007fd, Manning:2011ha}. The present experiments are consistent with these ideas: the irreversible rearrangements that could be activated by a given cyclic stress are exhausted by repeated applications, so that in the steady state a population of reversible rearrangements controls all deformation.

Our finding of microstructural reversibility suggests an unusual definition of yield strain $\gamma_y^\mathrm{micro}$ in these experiments: it is the largest strain amplitude at which the system can reach a steady state with reversible microstructure. This definition does not refer to any rheological measurement, only to the imposed stress or strain. By considering only the behavior at long times, we greatly reduce, if not eliminate, the dependence of this definition on material history. It is possible that the yielding transition is in fact a version of the critical transition observed by Cort\'e et al.~\cite{corte08}, with a transient duration that diverges at $\gamma_y^\mathrm{micro}$, warranting further experiments near $\gamma_y^\mathrm{micro}$.

The microstructural yield strain amplitude we obtain for our material, based on the data in Fig.~\ref{fig:relaxation}, is $\gamma_y^\mathrm{micro} \sim 0.03$, consistent with the simultaneously-measured rheological $\gamma_y^\mathrm{rheo} \sim 0.03$ in Fig.~\ref{fig:rheology}. In general it is not certain that these two $\gamma_y$ should be close. For example, the appearance of many reversible rearrangements could signal a partial loss of rigidity, so that $\gamma_y^\mathrm{micro} > \gamma_y^\mathrm{rheo}$. Similarly, it is possible that for $\gamma_0 < \gamma_y^\mathrm{rheo}$ experiments could observe significant irreversible changes after each cycle even in the steady state, so that $\gamma_y^\mathrm{micro} < \gamma_y^\mathrm{rheo}$. Conversely, H\'ebraud et al.~\cite{Hebraud:1997ef} found significant irreversibility in the steady state below $\gamma_y^\mathrm{rheo}$. It is thus notable that here, $\gamma_y^\mathrm{micro} = \gamma_y^\mathrm{rheo}$ within experimental resolution.

In this work, we have used our observation of microstructural reversibility to propose a way to experimentally determine a yield stress or strain for a jammed material without rheological measurements. Our results suggest this definition is compatible with bulk rheological definitions of yielding. Reversibility in cyclically-driven systems may thus have relevance to rheological yielding. Our results point toward a clear way to think about yielding in some materials: the bulk yielding transition corresponds to the largest cyclic deformation that leaves microstructure unchanged, over all possible material histories.

We thank John Bassani, Valeria Garbin, Daeyeon Lee, Bum Jun Park, Kate Stebe, Ye Xu, and Arjun Yodh. This work was supported by the Penn NSF MRSEC (DMR-1120901) and the Army Research Office (W911NF-11-1-0488).

\footnotesize{
\bibliography{references-nourl} 
\bibliographystyle{rsc} }

\end{document}